\documentclass[twocolumn,showpacs,preprintnumbers,amsmath,amssymb,superscriptaddress,10pt]{revtex4}
\usepackage{amssymb}
\usepackage{graphicx,color}
\usepackage{bm}

\begin{document}

\title{Elliptic and Triangular Flow and their Correlation in Ultrarelativistic
High Multiplicity Proton Proton Collisions at 14 TeV}

\author{Wei-Tian Deng}
\affiliation{Frankfurt Institute for Advanced Studies (FIAS),
Ruth-Moufang-Strasse 1, 60438 Frankfurt am Main, Germany}

\author{Zhe Xu}
\affiliation{Frankfurt Institute for Advanced Studies (FIAS),
Ruth-Moufang-Strasse 1, 60438 Frankfurt am Main, Germany}
\affiliation{Institut f\"ur Theoretische Physik, Johann Wolfgang Goethe-Universit\"at Frankfurt, Max-von-Laue-Strasse 1, 60438 Frankfurt am Main, Germany}
\affiliation{Physics Department, Tsinghua University, Beijing 100084, China}

\author{Carsten Greiner}
\affiliation{Institut f\"ur Theoretische Physik, Johann Wolfgang Goethe-Universit\"at Frankfurt, Max-von-Laue-Strasse 1, 60438 Frankfurt am Main, Germany}

\begin{abstract}
The spatial configuration of initial partons in high-multiplicity 
proton-proton scatterings at 14 TeV is assumed as three randomly positioned
``hot spots''. The parton momentum distribution in the hot spots is calculated
by HIJING2.0 with some modifications.
This initial condition causes not only large eccentricity $\epsilon_2$
but also triangularity $\epsilon_3$ and the correlation of $\epsilon_2-\epsilon_3$ 
event-plane angles. The final elliptic flow $v_2$, triangular flow $v_3$,
and the correlation of $v_2-v_3$ event-plane angles are calculated by 
using the parton cascade model BAMPS to simulate the space-time parton evolution. 
Our results show that the $v_2-v_3$ correlation is different from that of 
$\epsilon_2-\epsilon_3$. This finding indicates that translations of different Fourier 
components of the initial spatial asymmetry to the final flow
components are not independent. A dynamical correlation between the elliptic
and triangular flow appears during the collective expansion.
\end{abstract}


\maketitle

\section{Introduction}
A strong interest in high multiplicity events in ultrarelativistic
proton-proton collisions has arisen recently, since a near side ``ridge''
has been found in such events by the CMS collaboration at the Large Hadron
Collider (LHC) at $\sqrt{s}=7$ TeV \cite{Khachatryan:2010gv}. 
This phenomenon is very similar to that observed in Au-Au collisions 
at the Relativistic Heavy Ion Collider (RHIC) at $\sqrt{s_{NN}}=0.2$ TeV
\cite{Adams:2005ph,:2008cqb,Wenger:2008ts}. Because the quark-gluon 
plasma (QGP) created at RHIC behaves like a nearly perfect 
fluid \cite{Huovinen:2001cy}, one may
suppose that a similar hydrodynamic behaviour has appeared in high multiplicity
p-p collisions at LHC \cite{Werner:2010ss}.
Methods developed to investigate QGP at RHIC can be used to predict
new phenomena in p-p collisions at LHC. In this work
we are concentrating on the elliptic and triangular flow induced by
the initial eccentricity and triangularity in ultrarelativistic
high multiplicity p-p collisions at the highest LHC energy.
 
The elliptic flow parameter $v_2$ is the best experimental observable
determining the strength of the hydrodynamic collectivity
\cite{Arsene:2004fa,Back:2004je,Adams:2005dq,Adcox:2004mh}. 
In the presence of strong interactions
of system constituents, $v_2$ is obtained by the translation
of the spatial asymmetry of initially produced matter into the final 
particle angular distribution \cite{Ollitrault:1992bk,Voloshin:1994mz}. 
There are two ways to make a spatial asymmetry in nucleus-nucleus 
collisions. One is the geometric overlap in non-central collisions. 
This has been adopted to p-p collisions at LHC, because a proton also
has its extension though small. 
However, the predictions following the geometric overlap-eccentricity
$\epsilon_2$ showed small values of the elliptic flow parameter 
$v_2\approx 3\%$ in minimum bias, either in hydrodynamic 
calculations \cite{Luzum:2009sb,Prasad:2009bx,Ortona:2009yc} or in the 
$\epsilon_2-v_2$ scaling \cite{Cunqueiro:2008uu,d'Enterria:2010hd}.
The latter is assumed to be the same as that proposed for nucleus-nucleus
collisions at RHIC \cite{Bhalerao:2005mm,Drescher:2007cd}.
In the geometric picture non-central
collisions provide large initial eccentricity, which is the necessary
condition for large elliptic flow. However, the particle
multiplicity in such collisions in the case of p-p scatterings
is rather low. Nonflow effects can be hardly eliminated.

The second source of a spatial asymmetry comes from statistical density
fluctuations of the initially produced matter
on the event-by-event basis \cite{Qiu:2011iv}. 
This is the reason for the nonvanishing $v_2$ (and also higher harmonics)
in central nucleus-nucleus collisions at LHC \cite{:2011vk}.
Due to the much smaller volume of a proton compared with a Au nucleus, 
it is natural to consider fluctuations in central p-p
collisions at the LHC energy, which can provide both initial eccentricity and
high multiplicity. A few hot spots or flux tubes may be excited
in a p-p collision and lead to measurable elliptic flow at LHC
\cite{Bozek:2009dt,CasalderreySolana:2009uk,Chaudhuri:2009yp,Zhou:2010ij}. 
Moreover, parton evolutions and multiple scatterings 
\cite{Pierog:2010wa,Avsar:2010rf} can also generate large
event-by-event fluctuations.

Unlike the smooth initial distribution that generates
only even-order Fourier components in the momentum angular distribution,
initial fluctuations lead to nonvanishing odd-order components,
which are shown to contribute the azimuthal correlations observed at RHIC
\cite{Takahashi:2009na,Sorensen:2010zq,Alver:2010gr,Xu:2010du}.
In particular, the triangular flow $v_3$ from the initial triangularity 
$\epsilon_3$ is intensively studied in the recent works
\cite{Alver:2010dn,Petersen:2010cw,Qin:2010pf,Schenke:2010rr,Teaney:2010vd,Ma:2010dv,Petersen:2011fp}.
In central Au+Au collisions $v_3$ is as large as $v_2$ \cite{Ma:2010dv}.

For p-p collisions at LHC we suggest that $v_3$ is
as important as $v_2$ in high multiplicity events. Different from Au+Au
collisions at RHIC, where calculations \cite{Petersen:2010cw,Qiu:2011iv}
indicate no correlations between the initial event-plan angles of 
$\epsilon_2$ and 
$\epsilon_3$ and between the final event-plan angles of $v_2$ and $v_3$,
such correlations could exist in the p-p collisions at LHC
due to the smaller number of hot spots. Measurements on the flow correlations
will shed light on details in the context of collective flow
phenomena. In this work we calculate
$\epsilon_2$, $\epsilon_3$, $v_2$, $v_3$, and their correlations.
The event-by-event generation of the parton initial conditions is an 
implementation of the hot spots scenario \cite{CasalderreySolana:2009uk} 
by using the recent version of HIJING2.0 
\cite{Wang:1991hta,Deng:2010mv,Deng:2010xg}.
The dynamical space-time evolution is calculated by using
the parton cascade BAMPS \cite{Xu:2004mz}.

\section{Initial Conditions}
\label{init}
HIJING \cite{Wang:1991hta,Deng:2010mv,Deng:2010xg} is a Monte-Carlo event 
generator for hadron productions in high energy nucleon-nucleon, 
nucleon-nucleus, and 
nucleus-nucleus collisions. It is essentially a two-component model, which 
describes the production of hard parton jets and the soft interaction
between nucleon remnants. While the hard jets production can be calculated by 
perturbative QCD (pQCD), nucleon remnants interact via soft 
gluon exchanges described by the string model \cite{Sjostrand:1987su}. 
The produced hard jet pairs and the two excited remnants are treated as
independent strings, which fragment to resonances that decay to final
hadrons. The predictions using the updated HIJING2.0 \cite{Deng:2010mv}
are in good agreements with the recently measured hadron spectra at LHC
in p+p collisions at $\sqrt{s}=0.9$, 2.36 and 7 TeV \cite{:2009dt,Aamodt:2010pp,Aamodt:2010my,Khachatryan:2010xs,Khachatryan:2010us}, 
and central Pb+Pb collisions at $\sqrt{s_{NN}}=2.76$ TeV \cite{Aamodt:2010pb}.

\begin{figure}[h]
 \centering
 \includegraphics[width=0.45\textwidth]{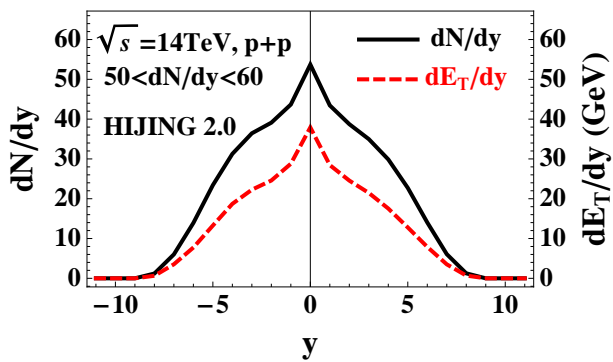}
 \caption{(Color online) The rapidity distribution of multiplicity and
transverse energy of hadrons produced in high multiplicity p+p collisions
at 14 TeV, calculated by HIJING2.0 \cite{Deng:2010mv}.}
 \label{fig:hadron_dndy}
\end{figure}

Using HIJING2.0 we calculate the hadron multiplicity for p+p collisions at 
14 TeV and find that the total hadron multiplicity $dN/dy$ at $y=0$ has
a mean value of 10.4 and possesses a high-multiplicity tail reaching the 
abundance of semi-peripheral Cu+Cu collisions at $\sqrt{s_{NN}}=62.4$ GeV 
at RHIC \cite{Alver:2006wh}. In this work we are interested in the events in
the window of $50 < dN/dy(y=0) < 60$. In these events the probability
to produce three strings (one is formed by the hard jet pairs and other two from
the excited remnants) is more than $80\%$.
We thus neglect, for simplicity, events without hard jets ($17\%$) and 
events containing more than one jet string ($3\%$). The rapidity distributions
of the hadron multiplicity and transverse energy in these selected events
are shown in Fig. \ref{fig:hadron_dndy}.
The peak at midrapidity is due to the fact that in the selected events,
the hard jets production and the multiple gluons exchange in the soft 
interaction mostly occur at midrapidity. 

\begin{figure}[h]
 \centering
 \includegraphics[width=0.45\textwidth]{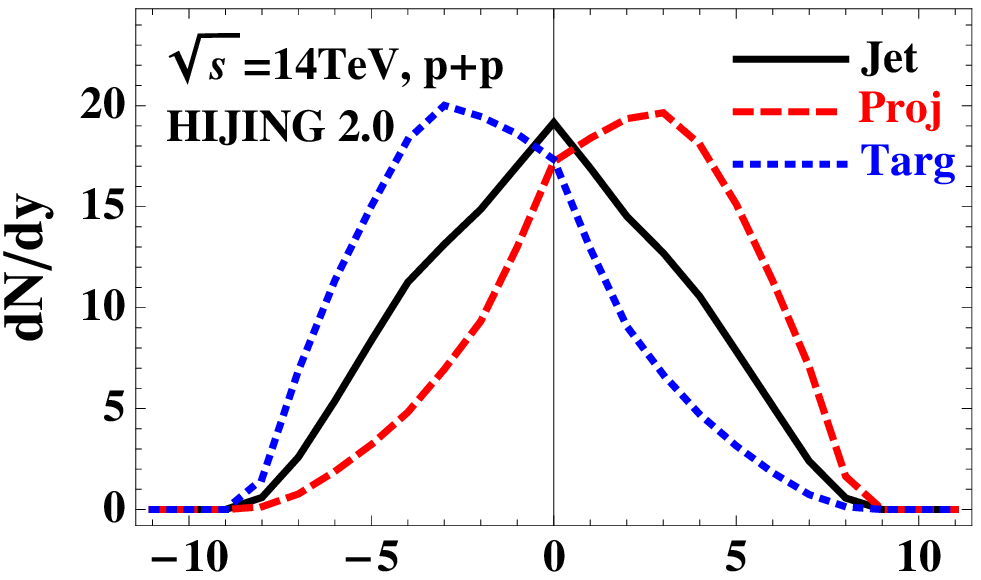}
 \includegraphics[width=0.45\textwidth]{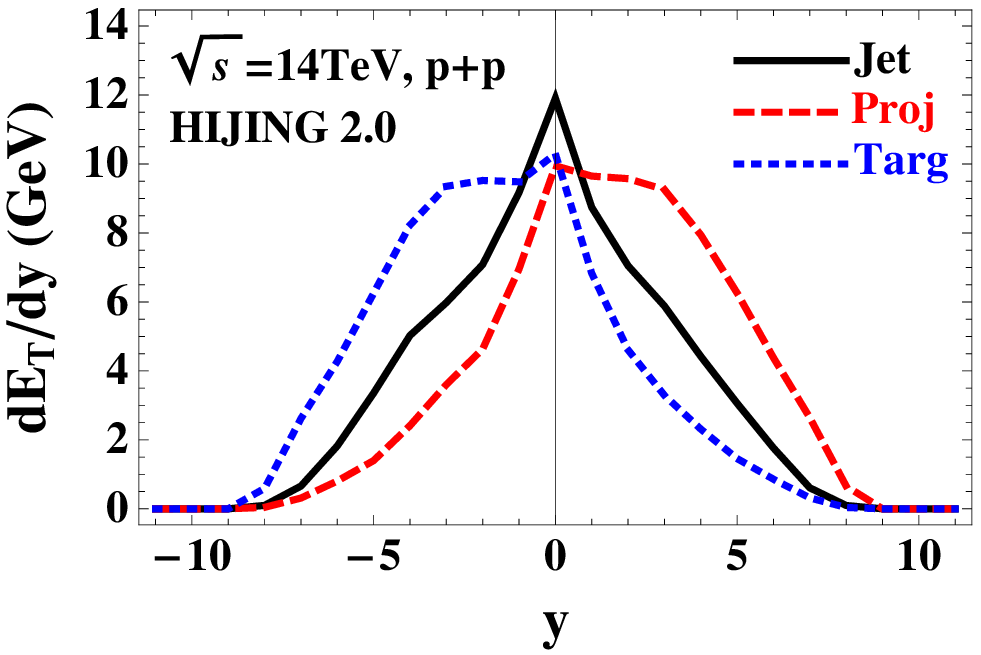}
 \caption{(Color online) The same as Fig. \ref{fig:hadron_dndy}, but on
the parton level from three strings. See details in text.}
 \label{fig:parton_dndy}
\end{figure}

To generate the initial condition on the parton level,
we turn off the resonance decays in HIJING and return to the representation of
resonances by quark-antiquark pairs or quark-diquark pairs according 
to the LUND string breaking \cite{Andersson:1986gw,Sjostrand:1987su}.
This approach is similar to the string melting
implemented in AMPT \cite{Lin:2001zk}, where final hadrons are converted
into partons. Figure \ref{fig:parton_dndy} shows the rapidity distribution
of parton number and transverse energy from jet, projectile, and target 
string, respectively.

Because the three strings break independently, we assume that partons
from each string build a hot spot. Due to the high collider energy all
partons are produced at $z=0$. The spatial distribution of the three
hot spots in the transverse plane follows the scenario proposed in 
Ref. \cite{CasalderreySolana:2009uk}. The center of the hot spots
is determined according to the proton spatial density \cite{Abe:1997xk}
\begin{equation}
 n_p(r)=\frac{n_0}{1+e^{(r-R_0)/d}} \,,
\end{equation}
where $n_0=0.17/fm^3$, $R_0=0.56$ fm and $d=0.112$ fm.
The spatial parton distribution in each hot spot is assumed to have
a Gaussian profile $e^{-r^2/r_0^2}$, where $r_0$ gives the size of
the hot spots.

Although $r_0$ of the remnant could be larger than that of the jet string,
we assume that $r_0$ is equal for each hot spot, which gives the 
largest effect on the initial asymmetry. $r_0$ is set to be 0.2 fm.
The smaller the $r_0$, the larger the $\epsilon_2$ and 
$\epsilon_3$ \cite{CasalderreySolana:2009uk}.

Translating particles into the frame where the average position is equal
zero, i.e., $\langle x \rangle=\langle y \rangle=0$, the harmonic components
$\epsilon_n$ of the spatial azimuthal distribution are defined 
as \cite{Petersen:2010cw}
\begin{eqnarray}
 \epsilon_n=\frac{\sqrt{\left\langle r^n\mathrm{cos}(n\phi) \right\rangle ^2 + \left\langle r^n\mathrm{sin}(n\phi) \right\rangle ^2}}{\left\langle r^n \right\rangle } \,,
 \label{eq:epsilon_n}
\end{eqnarray}
where ${r,\phi}$ are parton polar coordinates.
The corresponding initial event-plane angles are given by
\begin{equation}
 \Phi_n=\frac{1}{n}\mathrm{arctan}\frac{\left\langle r^n\mathrm{sin}(n\phi)\right\rangle }{\left\langle r^n\mathrm{cos}(n\phi)\right\rangle } \,.
 \label{eq:phi_n}
\end{equation}
These are the angles $\Phi=\Phi_n$ where 
$\langle -r^n \mathrm{cos}\,n(\phi-\Phi) \rangle/\langle r^n \rangle$
has the maximum, which is $\epsilon_n$ [see Eq. (\ref{eq:epsilon_n})].
For instance, for an ellipse shape with the short axis in the x direction
$\Phi_2=0$ and $\epsilon_2=\langle y^2-x^2 \rangle / \langle y^2+x^2 \rangle$.
Our definition of $\Phi_n$ differs from that from 
Ref. \cite{Petersen:2010cw} by $\pm \pi/n$.

\begin{figure}
 \centering
 \includegraphics[width=0.45\textwidth]{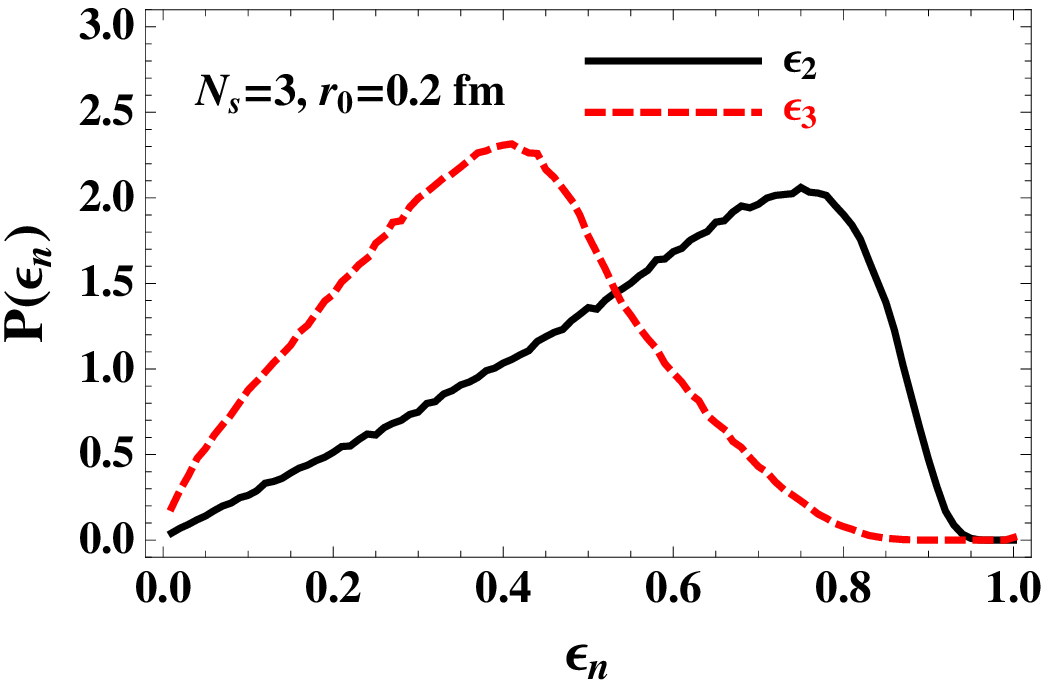}
 \includegraphics[width=0.44\textwidth]{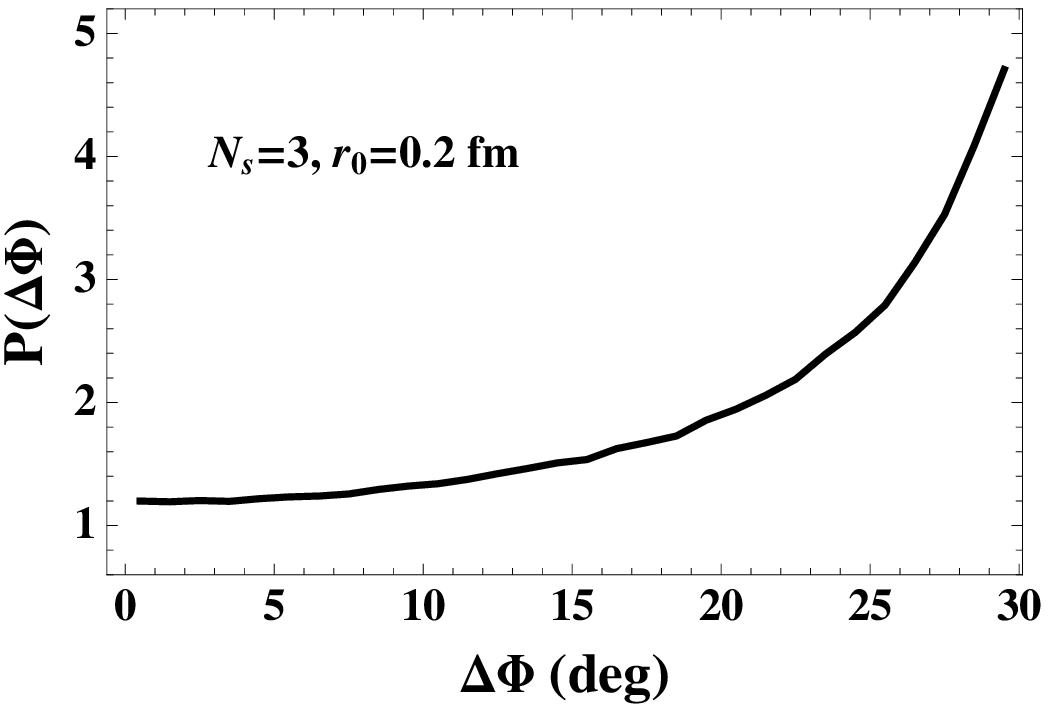}
 \caption{(Color online) Upper: The probability distribution of spatial
eccentricity and triangularity. Lower: Correlation (arbitrary unit) of the 
$\epsilon_2-\epsilon_3$ event-plane angles.}
 \label{fig:P_epsi}
\end{figure}

The upper panel of Fig. \ref{fig:P_epsi} shows the probabilities of
the eccentricity $\epsilon_2$ and the triangularity $\epsilon_3$
in the high-multiplicity p+p collisions at 14 TeV on the event-by-event basis.
The parton number in each hot spot is set to be 100. 
Using exact numbers from HIJING (about 160 in each hot spot taken over
all rapidities, see Fig.2) has tiny changes on the results. 100 partons
per hot spot are roughly the numbers within the rapidity range $|y|<2.5$,
which is a similar region covered in CMS \cite{Khachatryan:2010gv}.
Our results show that the most $\epsilon_3$ are not much smaller than
the most $\epsilon_2$, which indicates the possibility to measure
both $v_2$ and $v_3$ experimentally, if the assumptions of the initial
condition and the hydrodynamic transport are justified.

Due to the random nature of the initial condition, the distribution of 
$\Phi_2$ and $\Phi_3$ are uniform within the interval of $[-\pi/2,\pi/2]$
and $[-\pi/3,\pi/3]$, respectively. The lower panel of Fig. \ref{fig:P_epsi}
shows the probability distribution of the angle difference, 
$\Delta\Phi=|\Phi_2-\Phi_3|$, which indicates the correlation of the
two initial event-planes. Events with $\Delta\Phi=30$ degree are more probable.

For initial conditions with two hot spots instead of three, we expect 
vanishing $\epsilon_3$. With four or more hot spots both $\epsilon_3$
and $\epsilon_2$ become smaller and their event-plane angles will be
less correlated. Therefore, the largest correlation comes from events
with three hot spots.

When hot spots expand and overlap, collective flow components in higher 
order will be built up.
By analogy with $\epsilon_n$ flow coefficients $v_n$ are defined as
the harmonic components of the particle azimuthal distribution in momentum
space
\begin{equation}
 \frac{\mathrm{d}N}{\mathrm{d^2}p_T\mathrm{d}y}=\frac{1}{2\pi}\frac{dN}{p_T\mathrm{d}p_T\mathrm{d}y}\left[ 1+2\sum_n v_n\mathrm{cos}\,n(\psi-\Psi_n) \right]
\end{equation}
where
\begin{eqnarray} 
\label{eq:v_n}
&& v_n(p_T)= \langle \mathrm{cos}\,n(\psi-\Psi_n) \rangle \\
&& \Psi_n=\frac{1}{n}\mathrm{arctan}\frac{\langle \mathrm{sin}(n\psi)\rangle}{\langle \mathrm{cos}(n\psi)\rangle} \,.
\label{eq:psi_n}
\end{eqnarray}
$\Psi_n$ is the event plane angle \cite{Poskanzer:1998yz}, i.e., 
$\langle \mathrm{cos}\,n(\psi-\Psi) \rangle$ at $\Psi=\Psi_n$ has the 
maximum, which is $v_n$.

If the translations from $\epsilon_n$ to $v_n$ for all components
are completely independent, we will obtain $\Psi_n=\Phi_n$.
The strong correlation of $\Delta \Phi$ seen in Fig. \ref{fig:P_epsi} will 
be observed in $\Delta \Psi=|\Psi_2-\Psi_3|$ too. If such correlation
can be measured experimentally at LHC, this will be the evidence for the 
hot spot scenario of the initial condition and the hydrodynamic behaviour 
of the parton matter in high-multiplicity p+p collisions. In this work we
discuss this issue by simulating the parton collectivity within 
a microscopic manner.

\section{Elliptic, Triangular flow and their Correlation}
The space-time evolution of the partons is simulated by the parton cascade 
model BAMPS (Boltzmann Approach of MultiParton Scatterings) 
\cite{Xu:2004mz}, which solves the Boltzmann equation for on-shell
partons. For simplicity we regard partons, which stem from the string 
breaking (see Sec. \ref{init}), as identical massless Boltzmann
particles. The particle degeneracy factor is assumed to be the same as that
of gluons together with quarks with two flavours. This leads to the particle
number density $n_{eq}=(40/\pi^2)T^3$, where $T$ is the temperature,
if the system is in local thermal and chemical equilibrium.
Furthermore, we consider only elastic binary scatterings
and assume the isotropic distribution of the collision angle.

The Boltzmann equation applies for systems when the particle mean free path 
$\lambda_{mfp}=1/(n\sigma)$ is larger than the mean particle distance
$d=n^{-1/3}$, where $n$ is the local particle number density. 
For chosen constant $\lambda_{mfp}/d$ ratio as a global parameter
we determine the total cross sections $\sigma$ in local cells, which are 
used to simulate scatterings \cite{Xu:2004mz}.

To assess hydrodynamic behaviour of the partons considered in this work
we calculate the shear viscosity to the entropy density ratio $\eta/s$
for given $\lambda_{mfp}/d$ ratio.
The shear viscosity is proportional to the energy density and
the mean free path \cite{El:2011cp}, $\eta=(2/5)e\lambda_{mfp}$. 
Assuming local thermal equilibrium, the entropy density is 
$s=(4-\ln \lambda_p)n$, where $\lambda_p=n/n_{eq}$ denotes the particle
fugacity. We have then 
$\eta/s=0.752\lambda_{mfp}/d/(4-\ln \lambda_p)/\lambda_p^{1/3}$.
The relation $e=3nT$ is used in the last equation. The ratio
$\eta/s$ has a weak dependence on the fugacity. Therefore, we take
$\eta/s=0.188\lambda_{mfp}/d$, which is exact for $\lambda_p=1$,
as an estimate of
the $\eta/s$ value for systems out of equilibrium, such as the present case.
For choosing $\lambda_{mfp}/d=2$ for instance we obtain 
$\eta/s\approx 0.376$.

Further model parameters are set as follows:
The initial time for starting BAMPS is $0.1$ fm/c. Before that
time partons propagate freely. The cell length in the transverse plane is
$\Delta x=\Delta y=0.02$ fm, while the longitudinal cell size is 
$\Delta \eta \approx 0.1$, expressed in the space-time rapidity 
$\eta=0.5 \ln [(t+z)/(t-z)]$. The test particle method \cite{Xu:2004mz} 
is used to enhance the numerical accuracy. For that the parton density
is amplified by a factor of 3000. Cross sections are reduced by the
same factor to keep the mean free path unchanged. Parton scatterings
stop when the local energy density is lower than $1 \rm{GeV/fm^3}$.
This mimics the phase transition, which has to be implemented in BAMPS
in the future. For the event-by-event analysis we compute ten thousands 
runs.

\begin{figure}[h]
 \centering
 \includegraphics[width=0.45\textwidth]{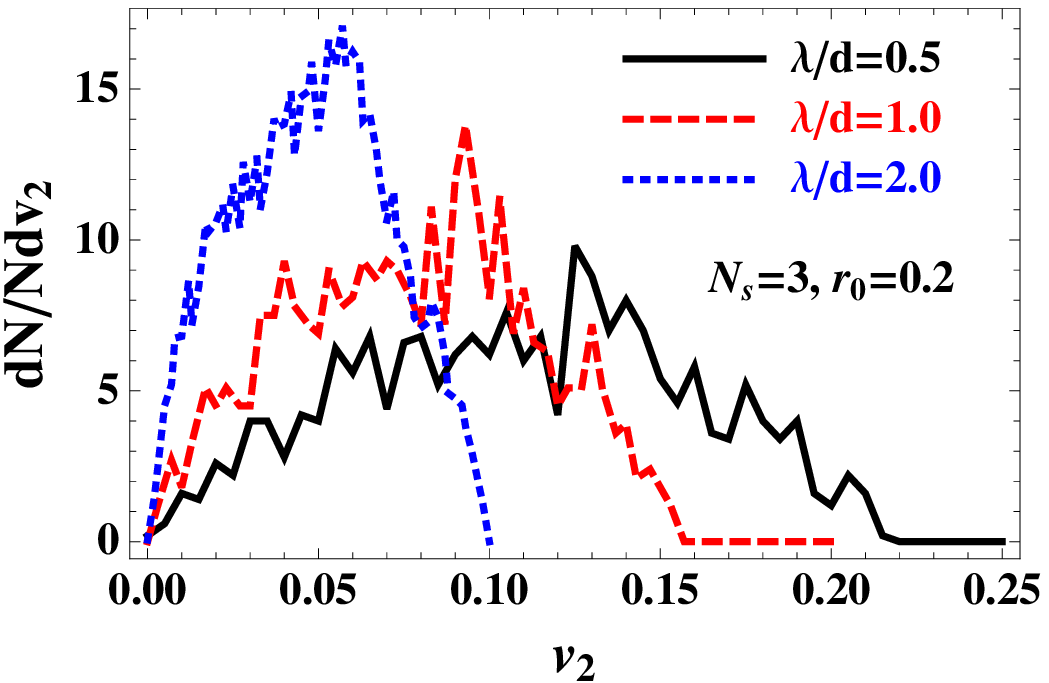}
 \includegraphics[width=0.45\textwidth]{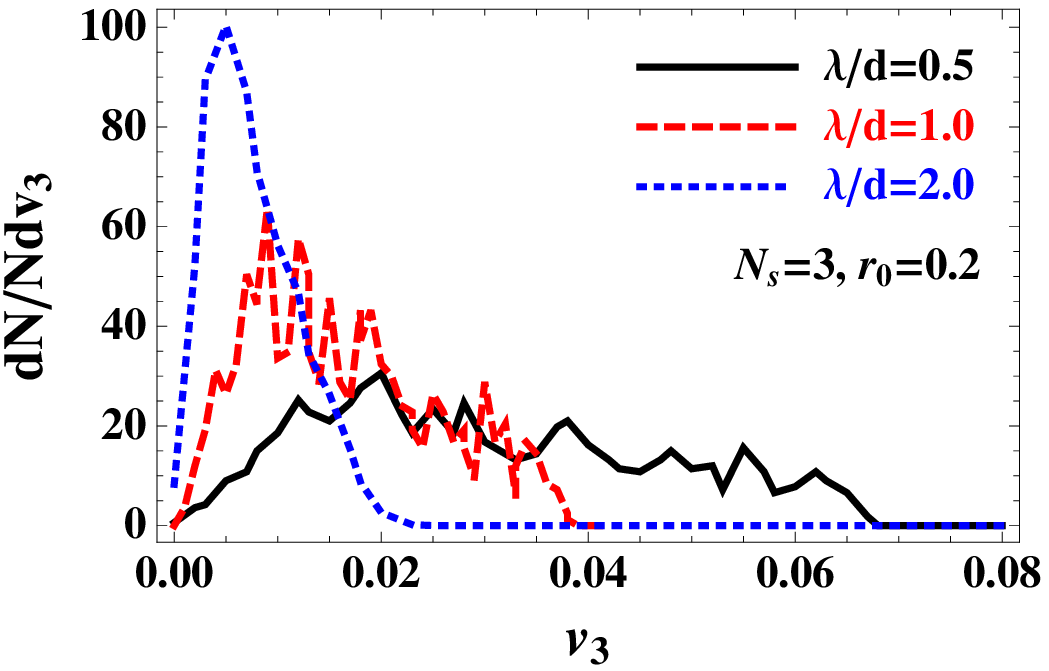}
 \caption{(Color online) Event-by-event distribution of $v_2$ and $v_3$.}
 \label{fig:P_v}
\end{figure}

Figure \ref{fig:P_v} shows the event-by-event distribution of the 
$p_T$-averaged $v_2$ and $v_3$.
For $\lambda_{mfp}/d=2$ $v_2$ has a broad distribution between 0 and 0.1 
with the maximum at 0.06,
which is comparable with the values at RHIC. $v_3$'s distribution 
is narrow and centered at 0.01. For smaller $\lambda_{mfp}/d$
ratios the viscous effect becomes smaller and the collective flow becomes
stronger. The value of $v_2$ and $v_3$ can reach 0.2 and 0.07, respectively,
for $\lambda_{mfp}/d=0.5$,  which corresponds to $\eta/s\approx 0.094$.
Therefore, if the parton matter in
high-multiplicity p+p events at LHC has a small $\eta/s$ ratio, 
both $v_2$ and $v_3$ are measurable quantities. 
Although the Boltzmann equation is not strictly valid for systems with
$\lambda_{mfp}/d < 1$, its solution agrees well with results from 
hydrodynamic calculations with corresponding $\eta/s$ 
ratio \cite{Bouras:2009nn}.

\begin{figure}[ht]
 \centering
 \includegraphics[width=0.45\textwidth]{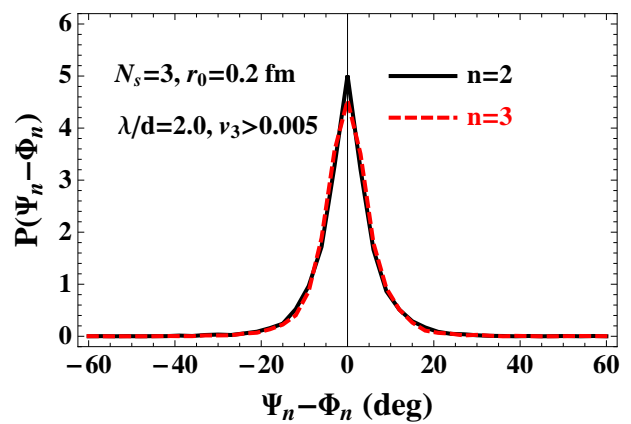}
 \caption{(Color online) Distributions (arbitrary unit) of the difference
between the initial event-plane angle $\Phi_n$ and corresponding final
one $\Psi_n$.}
 \label{fig:P_deltaPhi_cut_transfer}
\end{figure}

We are more interested in the translation from the initial spatial asymmetry
to the final collective flow. In Fig. \ref{fig:P_deltaPhi_cut_transfer}
we show the distributions of the difference between $\Psi_n$ and $\Phi_n$
for $n=2,3$ [see the definitions in Eqs. (\ref{eq:phi_n}) and
(\ref{eq:psi_n})].
The distributions of $n=2,3$ are almost the same. They peak at zero and
have a form looking like the Dirac delta function. It seems that
translations from $\epsilon_2$ to $v_2$ and from $\epsilon_3$ to $v_3$
take place independently.

\begin{figure}[ht]
 \centering
 \includegraphics[width=0.45\textwidth]{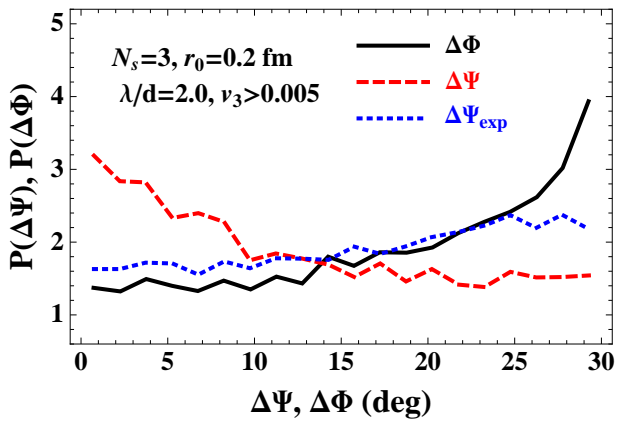}
 \caption{(Color online) Correlations (arbitrary unit) of initial 
$\epsilon_2-\epsilon_3$ and final $v_2-v_3$ event-plane angles. The
dotted curve is obtained by independent samplings of $\Psi_2$ and $\Psi_3$
according to Fig. \ref{fig:P_deltaPhi_cut_transfer}. Events with
$v_3>0.005$ are selected.}
 \label{fig:P_deltaPhi_v3G0.005}
\end{figure}

However, the event-plane correlations present a different picture, which is
given in Fig. \ref{fig:P_deltaPhi_v3G0.005}.
We have selected events with $v_3>0.005$, because in events with $v_3<0.005$
$\Psi_3$ (possibly also $\Phi_3$ due to tiny $\epsilon_3$) is a random 
number within $[-\pi/3,\pi/3]$ and has no correlation with the initial 
$\Phi_3$. The solid curve shows the angular correlations of the initial 
$\epsilon_2-\epsilon_3$ event-planes, which is almost the same as
the one demonstrated with a simpler initial condition, seen in
Fig. \ref{fig:P_epsi}. The dashed curve shows the corresponding
correlation of the $v_2-v_3$ event-planes after parton cascade simulations. 
Surprisingly, the $v_2-v_3$ event-plane correlation function has a maximum
at zero degree and, thus,
is totally different from the $\epsilon_2-\epsilon_3$ one.
With the same event-by-event initial conditions but without parton cascade
simulations we sample $\Psi_2$ and $\Psi_3$ independently according to 
Fig. \ref{fig:P_deltaPhi_cut_transfer}. The result is plotted by the dotted 
curve in Fig. \ref{fig:P_deltaPhi_v3G0.005}, which shows, as expected,
the same trend as the initial angular correlation, although the correlation
is weaker due to the width in the distributions seen in 
Fig. \ref{fig:P_deltaPhi_cut_transfer}. Because the independent
samplings of $\Psi_2$ and $\Psi_3$ does not reproduce the $\Psi_2-\Psi_3$
correlation (dashed curve) in the parton cascade calculations, 
we conclude that elliptic and triangular flow are correlated during 
the dynamical expansion. This dynamical correlation seems to rotate different
event-planes (30 degree in $\Delta \Phi$) to a unified event-plane
(0 degree in $\Delta \Psi$).

\begin{figure}[ht]
 \centering
 \includegraphics[width=0.45\textwidth]{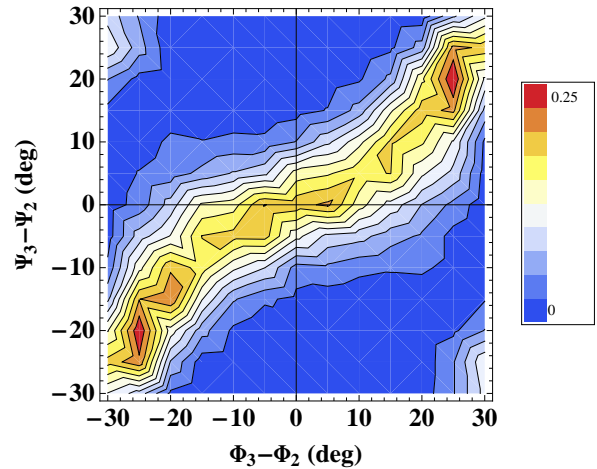}
 \caption{(Color online) Contour plot 
$d^2N/d(\Phi_3-\Phi_2)/d(\Psi_3-\Psi_2)$ (arbitrary unit).}
 \label{fig:contour1}
\end{figure}

\begin{figure}[ht]
 \centering
 \includegraphics[width=0.45\textwidth]{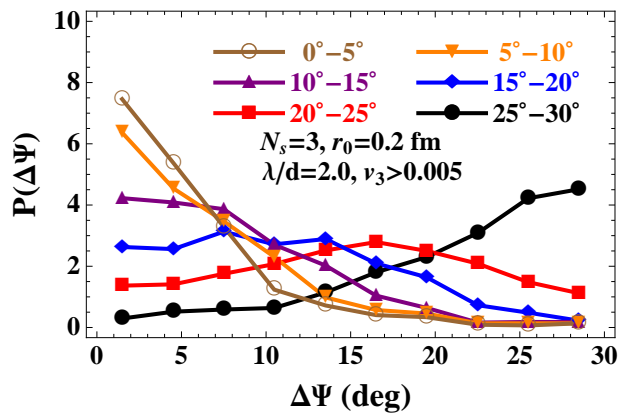}
 \caption{(Color online) Correlations (arbitrary unit) of final $v_2-v_3$
event-plane angles within the separate bins of the initial correlation
angles, $|\Phi_2-\Phi_3|=$0-5, 5-10, 10-15, 15-20, 20-25, and 25-30 degree.}
 \label{fig:many_deltaPhi_v3G0.005}
\end{figure}

To convince ourselves of the finding, we show the contour plot 
$d^2N/d(\Phi_3-\Phi_2)/d(\Psi_3-\Psi_2)$ in Fig. \ref{fig:contour1}.
Integral over $\Psi_3-\Psi_2$ gives the solid curve in 
Fig. \ref{fig:P_deltaPhi_v3G0.005}, while integral over $\Phi_3-\Phi_2$
gives the dashed curve in Fig. \ref{fig:P_deltaPhi_v3G0.005}.
The difference between the solid and the dashed curve in 
Fig. \ref{fig:P_deltaPhi_v3G0.005} is reflected in the asymmetry 
in Fig. \ref{fig:contour1} along the plane $\Psi_3-\Psi_2=\Phi_3-\Phi_2$.
At a fixed $\Phi_3-\Phi_2$, $\Psi_3-\Psi_2$ has a broad distribution with
a center moving toward $\Psi_3-\Psi_2=0$. This is better observed in 
Fig. \ref{fig:many_deltaPhi_v3G0.005}, where the final $|\Psi_2-\Psi_3|$
correlations from events within separate bins of the initial correlation
angles, $|\Phi_2-\Phi_3|=$0-5, 5-10, 10-15, 15-20, 20-25, and 25-30 degree,
are shown.
We clearly see strong broadening of all the distributions toward zero degree.
For instance, we choose the curve denoted by $10-15$ degree.
When assuming independent translations from $\epsilon_n$ to $v_n$,
this curve is expected to peak at $10-15$ degree. However,
we realize its maximum at zero degree.

\begin{figure}[ht]
 \centering
 \includegraphics[width=0.45\textwidth]{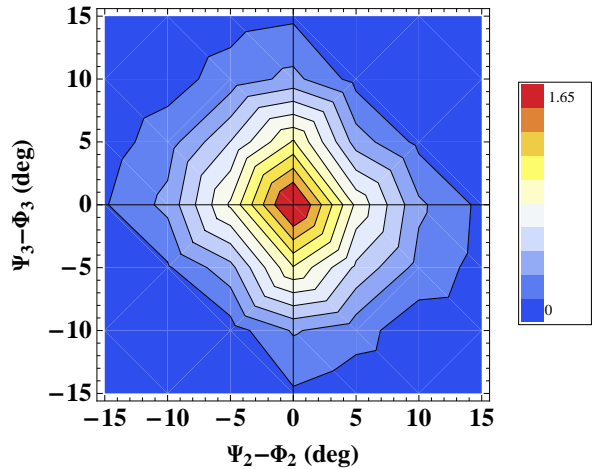}
 \caption{(Color online) Contour plot 
$d^2N/d(\Psi_2-\Phi_2)/d(\Psi_3-\Phi_3)$ (arbitrary unit).}
 \label{fig:contour2}
\end{figure}

Figure \ref{fig:contour2} shows another contour plot 
$d^2N/d(\Psi_2-\Phi_2)/d(\Psi_3-\Phi_3)$.
Integral over $\Psi_3-\Phi_3$ (or $\Psi_2-\Phi_2$) gives the solid
(or dashed) curve in Fig. \ref{fig:P_deltaPhi_cut_transfer}. 
If $\Psi_2-\Phi_2$ and $\Psi_3-\Phi_3$ are independent, the contour
structure should be symmetric along the planes $\Psi_2-\Phi_2=0$
and $\Psi_3-\Phi_3=0$. It is not the case. Assuming $\Phi_2=0$ and
$\Phi_3=30$ degree, (which are more favored,) and choosing a $\Psi_2$ with
$\Psi_2-\Phi_2>0$, we see from the contour plot that events with 
$\Psi_3-\Phi_3<0$ are more favored. This indicates that the angle
between $\Psi_2$ and $\Psi_3$ is smaller than 30 degree and the 
two event-planes rotate toward each other during the dynamical expansion.

We have to note that our conclusion on the dynamical correlation
between $v_2$ and $v_3$ needs further verifications, because the 
fluctuating initial configuration of hot spots may affect the final 
event-plane angular correlation.
It is worthwhile to study this issue for a smooth initial condition
with few components $\epsilon_n$ and given initial event-plane
angular correlation. We leave this as a task for a future 
investigation.

\section{Conclusions}
In this work we have calculated the elliptic and triangular flow
in high-multiplicity proton-proton collisions at the LHC energy
14 TeV. The reason for the measurable flows is the assumed initial
fluctuation in the hot spot scenario.
The motivation of this work was to find final $v_2-v_3$ event-plane
correlation expected by the initial $\epsilon_2-\epsilon_3$ event-plane
correlation. The latter exists for the assumed initial condition with
three statistically distributed hot spots originating from three 
independent fragmenting strings in p+p collisions modelled by HIJING. 
The results obtained by using parton transport model
BAMPS showed the largest $v_2-v_3$ event-plane angular correlation at
zero degree, which is the opposite to the expectation at 30 degree.
This observation indicates a dynamical correlation between elliptic
and triangular flow during the expansion. Their event-planes rotate 
toward each other. If so, any initial correlations will be washed out 
and it is more difficult to extract initial conditions from the flow 
observations. On the other hand, because we do not expect initial
event-plane angular correlations in
nucleus-nucleus collisions at RHIC and LHC, measurements on the final
harmonic flow event-plane angular correlations in these experiments would
confirm our conclusion, if data favor zero-degree correlation of $v_2$,
$v_3$ event-plane angles.
The correlation of different flow components is a new finding and
needs further verifications in the future.

\section*{Acknowledgments}
WTD would like to acknowledge J. Casalderrey-Solana, Y.-K. Song and 
X.-N. Wang for useful discussions. ZX thanks S. Esumi for his fruitful 
suggestions. The BAMPS simulations were performed 
at the Center for Scientific Computing of Goethe University. 
This work was financially supported by the Helmholtz International Center
for FAIR within the framework of the LOEWE program launched by the State
of Hesse.

\end{document}